\documentstyle[psfig]{mn}

\title[$UBVRI$ photopolarimetry of V1309 Ori]
{$UBVRI$ photopolarimetry of the long period eclipsing 
AM Herculis binary V1309 Ori.}

\author[S. Katajainen et al.]
 {S.~Katajainen,$^1$
   V.~Piirola,$^1$    
   G.~Ramsay,$^2$    
   F.~Scaltriti,$^3$\thanks{Visiting Astronomer, 
 Complejo Astronomico El Leoncito operated 
under agreement between the Consejo Nacional de Investigaciones y 
Tecnicas de la Republica Argentina and the National Universities of La 
Plata, Cordoba and San Juan}      
H.J.~Lehto,$^{1,4}$     
M.~Cropper,$^2$   
 \newauthor
M. K.~Harrop-Allin,$^2$ 
E.~Anderlucci,$^{3,\star}$ and P.~Hakala$^1$\\ 
$^1$Tuorla Observatory, University of Turku, FIN-21500, 
   Piikki\"o, Finland\\
$^2$Mullard Space Science Laboratory, University College London, 
  Holmbury St Mary, Dorking, Surrey RH5 6NT, UK\\
$^3$Osservatorio Astronomico di Torino, I-10025 Pino Torinese, 
 Italy\\
$^4$Department of Physics, University of Turku, FIN-21400, Finland\\}
     
\date{Received:}

\begin{document}
\newcommand{\Msun} {$M_{\odot}$}

\maketitle

\begin{abstract}

We report simultaneous $UBVRI$ photo-polarimetric observations of the
long period (7.98 h) AM Her binary V1309 Ori.  The length and shape of
the eclipse ingress and egress varies from night to night. We suggest
this is due to the variation in the brightness of the accretion stream. By
comparing the phases of circular polarization zero-crossovers with
previous observations, we confirm that V1309 Ori is well synchronized,
and find an upper limit of 0.002 percent for the difference between
the spin and orbital periods. We model the polarimetry data using a
model consisting of two cyclotron emission regions at almost
diametrically opposite locations, and centered at colatitude
$\beta=35^{\circ}$ and $\beta=145^{\circ}$ on the surface of the white
dwarf. We also present archive X-ray observations which show that the
negatively polarised accretion region is X-ray bright.

\end{abstract}

\begin{keywords}
stars:individual: V1309 Ori -- stars: magnetic fields -- novae, cataclysmic
variables -- white dwarfs : stars
\end{keywords}

\section{Introduction}

The X-ray source RX J0515.6+0105 (V1309 Ori) was preliminarily
identified in the {\sl ROSAT} All Sky Survey as a Cataclysmic Variable
(CV) with an orbital period of $\sim$8 hrs \cite{b2}.
Garnavich et al. \shortcite{b12} showed that V1309 Ori has properties which are
typical of magnetic CVs (mCVs). They also found a deep total eclipse
in the light curves. Pointed X-ray observations made using {\sl ROSAT}
\cite{b45} showed extremely variable X-ray emission
in short bursts (up to few seconds), suggesting inhomogeneous
accretion of dense blobs.  Detection of variable circular polarization
in white light by Buckley \& Shafter \shortcite{b4} confirmed the
classification of V1309 Ori as an AM Her star (synchronised mCVs).

The orbital period of V1309~Ori 7.98 h is $\sim$3 hours longer than in
any other known AM Her system. Only three AM Hers are known to have
periods over 4 hours (RX J1313--32: 4.25 h; AI Tri: 4.59 h; V895 Cen:
4.77 h: Ritter \& Kolb 1998). The exceptionally long orbital period
indicates that the separation between the primary and the secondary is
large for mCV.  According to Patterson's \shortcite{b25} scaling law, magnetic
field field strengths up to 150 -- 730 MG should be required to
synchronize the spin of the white dwarf with the rotation of the
binary system, assuming a mass for the white dwarf $M_{\rm
WD}$=0.6-1.0 M$_{\odot}$ \cite{b35}. However, spectroscopic
studies have given significantly lower magnetic field values: 33 -- 55
MG Garnavich et al. \shortcite{b12}; 61 MG Shafter et al. \shortcite{b35}; 
\( < 70 \) MG de Martino et al. \shortcite{b8}.
Frank, Lasota \& Chanmugam \shortcite{b10} have proposed that these low
magnetic field values can be understood in terms of the standard
evolutionary model, if the system is in a low accretion state for a
long enough time or if the magnetic field of the secondary is strong
enough (\( > 1\)kG) to maintain synchronism.  For these reasons, V1309
Ori is one of the most important objects for studying synchronization
and accretion processes in mCVs and their evolution. In this paper we
present the results and analysis of our polarimetric, photometric and
spectroscopic observations of this system.

\begin{table*}
\centering
\begin{minipage}{140mm}
\caption{The observing Log for V1309 Ori}
\begin{tabular}{rrrrrrr}
\hline   
UT Date  & HJD Start & Filter(s) & Duration & Exp. Time & 
Telescope& Type of Observations \\
(at start)  &(2,440,000.0+)& & (hr) & (sec) & & \\
\hline 
Oct 7 1997 & 10728.5920 & UBVRI & 4.42 & 10 & NOT & photopol. (lin.)\\
Oct 8 1997 & 10729.5890 & UBVRI & 4.82 & 10 & NOT & photopol.(cir. \& lin.)\\
Oct 9 1997 & 10730.5875 & UBVRI & 4.65 & 10 & NOT & photopol.(cir. \& lin.)\\
\hline
Nov 24 1997 & 10777.4853 & UBVRI & 8.06 & 10 & NOT & photopol.(cir. \& lin.)\\
\hline
Dec 31 1997 & 10813.6973 & UBVRI & 0.17 & 10 & CASLEO & photopol. (lin.)\\
Jan 1 1998 & 10814.5737 & UBVRI & 3.17 & 10 & CASLEO & photopol. (lin.)\\
Jan 2 1998 & 10815.5954 & UBVRI & 2.85 & 10 & CASLEO & photopol. (lin.)\\
Jan 4 1998 & 10817.5781 & UBVRI & 2.37 & 10 & CASLEO & photopol. (lin.)\\
Jan 5 1998 & 10818.5980 & UBVRI & 3.01 & 10 & CASLEO & photopol. (lin.)\\
\hline
Dec 25 1998 & 11172.9840 & Blue$^{1}$ \& Red$^{2}$ & 4.84 & 300 & ANU &
 spectrosc.\\
Dec 26 1998 & 11174.0080 & Blue$^{1}$ \& Red$^{2}$ & 2.95 & 300 & ANU & 
spectrosc. \\
Dec 27 1998 & 11174.9310 & Blue$^{1}$ \& Red$^{2}$ & 6.29 & 300 & ANU & 
spectrosc.\\ 
\hline
\end{tabular}
\medskip 

NOT: 2.56 m Nordic Optical Telescope, Observatorio del Roque de los Muchachos,
La Palma, Spain; \\ 
CASLEO: 2.15 m telescope, Complejo 
Astronomico El Leoncito, province of San Juan, Argentina; \\
ANU: 2.30 m Australian National University telescope, 
Siding Spring Observatory (SSO), Australia; \\
Blue$^{1}$: wavelength coverage 3800\AA -- 5000\AA\\  
Red$^{2}$: wavelength coverage 6200\AA -- 7500\AA \\
\end{minipage}
\end{table*}

\section{Observations}

Photo-polarimetric observations were made at the NOT (Nordic Optical
Telescope, La Palma) using the Turpol-photo-polarimeter in October and
November 1997 (see table 1 for the log of observations).  Observations
were also made on December 1997 and January 1998 with the 2.15 m
CASLEO-telescope (Argentina) using the Turin-photo-polarimeter. This
instrument at CASLEO is almost identical to the Turpol at the NOT.
 
The polarimeters have four dichroic filters, splitting the light into
$UBVRI$ bands.  The time resolution for photometric data is 24 sec.
One polarization measurement consists of eight integrations and takes
$\sim$3~minutes. In the simultaneous circular \& linear mode with the
$\lambda$/4--retarder the efficiency for circular polarization is
about 70 per cent and for linear polarization about 50 per cent. With
the $\lambda$/2--retarder $\sim$ 100 percent efficiency for linear
polarization is achieved. The seeing was between 0.6 and 1.5 arcsec
during all the nights at the NOT and the 7.5-arcsec diaphragm was
used. At the CASLEO, average seeing was between 2 and 3 arcsec and the
diaphragm 11 arcsec.
  
Sky background polarization was eliminated by using a calcite plate as
a beam splitter. Sky intensity was measured at 15 -- 30 min
intervals.  Instrumental polarization and the zero-point of position
angle were determined from observations of standard stars 
BD+32$^{\circ}$3739, HD 204827, BD+64$^{\circ}$106, HD161056, and
HD155197 \cite{b32}. Photometric UBVRI standard stars
92282, 94242, 97351, 110340, and 114750 \cite{b20} were used to
calibrate the photometry.

Spectroscopic observations of V1309 Ori were carried out on 3 nights,
starting on 25 Dec 1998, using the ANU (Australian National
University) 2.3 m telescope at Siding Spring Observatory (SSO),
Australia. Spectra were obtained with the double beam spectrograph
using 600 lines/mm gratings. The effective wavelength coverage was
3800--5000 ${\rm \AA}$ and 6200--7500 ${\rm \AA}$ in the blue and red
arms of the spectrograph, respectively. The conditions were
photometric throughout the observations, and the effective resolution
as measured from the FWHM of the arc lines was 2.1 \AA. The exposure
time was 300 sec for both blue and red spectra. Wavelength calibration
of the flat-fielded and bias-subtracted two dimensional images was
performed using He-Ar arc spectra taken at various points during the
night.

The photometric ephemeris of Staude, Schwope \& Schwarz \shortcite{b42} 
(their equation (1)) is used to phase the data throughout the paper.
 
\section{Photometry}

\begin{figure}
\psfig{file=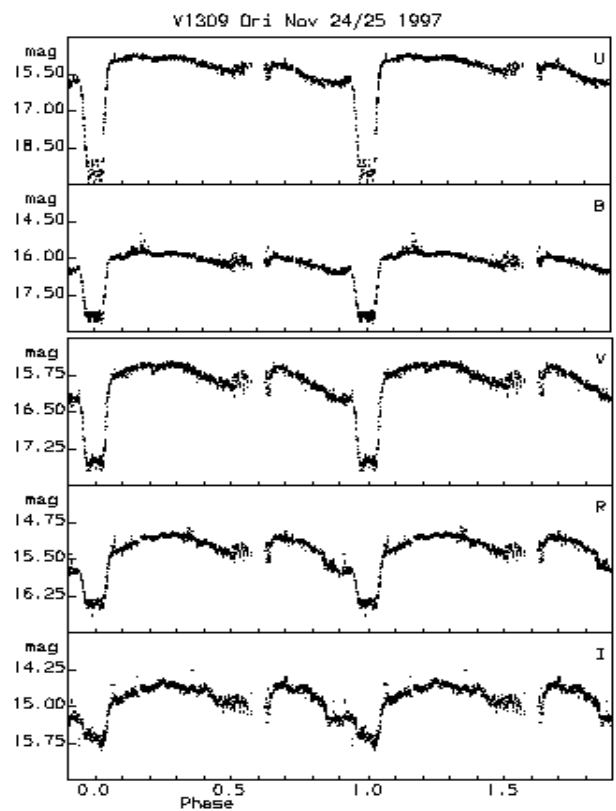,width=11cm,height=11cm}
\caption{Simultaneous $UBVRI$ light curves of V1309 Ori observed at
the NOT on November 24/25, 1997.  Each point presents a single
photometric measurement, with a 24 s time resolution. Data have been
plotted twice to clarify brightness variations over complete
orbital cycle. The short gap (15 min) near the phase $\Phi=0.6$ is
due to cloud. Note the different scale in the $UB$ and the $VRI$ bands.}
\label{ubvri}
\end{figure}

\begin{figure}
\psfig{file=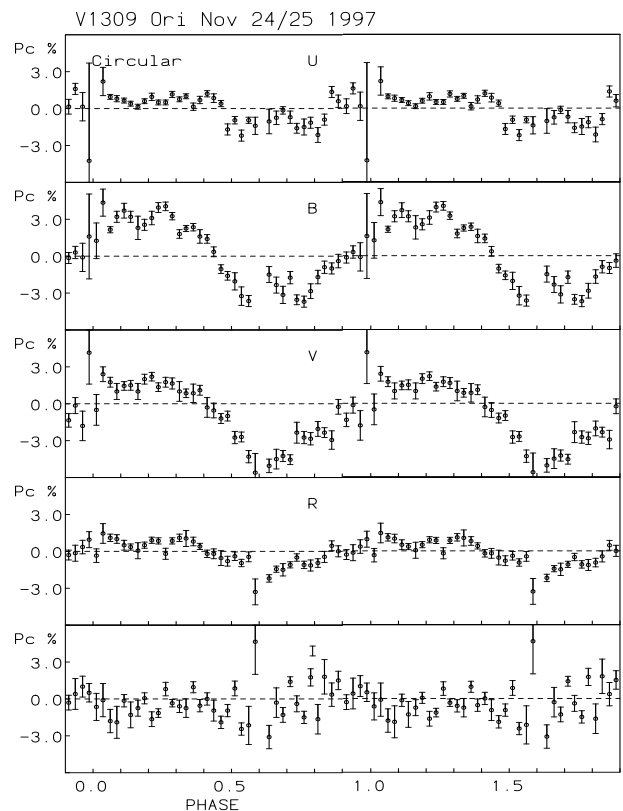,width=11cm,height=11cm}
\caption{$UBVRI$ circular polarization curves obtained on November 24/25, 
1997, simultaneously with the photometry shown in Figure 1.}
\label{circpol}
\end{figure}

\begin{figure}
\psfig{file=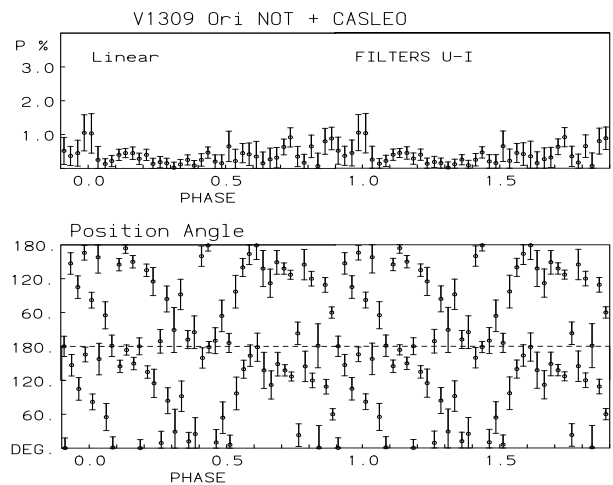,width=9cm,height=7cm}
\caption{Linear polarization and position angle curves for V1309 Ori. 
Data have been combined from 4 nights in October and November, 
1997 at the NOT, and on 5 nights in January 1998 at the CASLEO. 
Polarization has been computed by vectorially averaging individual 
observations in the $UBVRI$ bands into 40 phase bins.}
\label{linpol}
\end{figure}

During our observations V1309 Ori was in a high accretion state
($V$=16), as found also in earlier photometric studies (Garnavich et
al. 1994; Shafter et al. 1995).  In Figure \ref{ubvri} we present
simultaneous $UBVRI$ light curves covering nearly the full orbital
cycle, obtained on November 24/25, 1997, at the NOT. Our observations
made in October (NOT) 1997 and in January 1998 (CASLEO) do not cover
the complete orbital cycle (and are not shown here) show that V1309
Ori was in the same brightness level (V=16) as in November 1997. We
show the circular polarisation data in Figure \ref{circpol} and the
linear polarisation data in Figure \ref{linpol}.

The most distinctive feature of the intensity curves (Figure
\ref{ubvri}) is the well known deep eclipse which has a duration of
$\sim$0.1 orbital phase and also a strong colour dependence (up to
$\sim$4 mag in $U$, and less than 1 mag in $I$).  Flux variations
outside the eclipse are about 1.5 mag in the $U$, and about one
magnitude in the $BVRI$ bands. Light curves show two local brightness
maxima, at orbital phases $\Phi=0.2$ and $\Phi=0.7$.  There is a clear
asymmetry between these two maxima: in the $U$-band the first peak
near $\Phi=0.2$ is about 0.5 mag brighter than observed at $\Phi=0.7$,
whereas in the $B$- and $V$-bands this difference is about 0.3 mag and
in the $I$-band typically about 0.2 mag.

\begin{figure}
\psfig{file=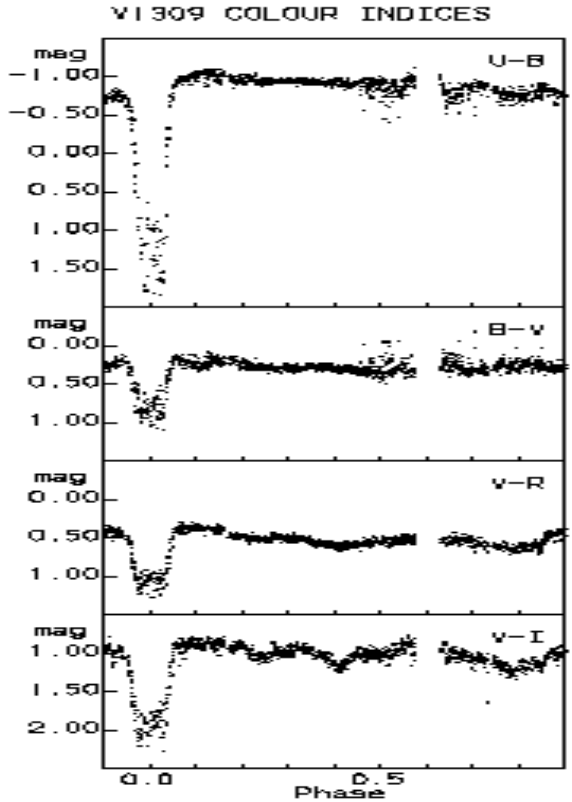,width=9cm,height=11cm}
\caption{Colour indices  $U-B$, $B-V$,
$V-R$, and $V-I$ of V1309 Ori (November 24/25, 1997), plotted 
over the orbital cycle.}
\label{colour}
\end{figure}

The colour indices in Figure \ref{colour} show V1309 Ori $bluer$ after
the eclipse, where the system is also brighter than before the eclipse
(Figure \ref{ubvri}).  Significant colour changes take place over the
whole orbital period, when different parts of the stream are viewed at
different angles. Eclipse of the secondary star by the stream also
contributes, noticeably in the $R$ and $I$ bands in the phase interval
0.4-0.7.

\subsection{The eclipse}

\begin{figure}
\psfig{file=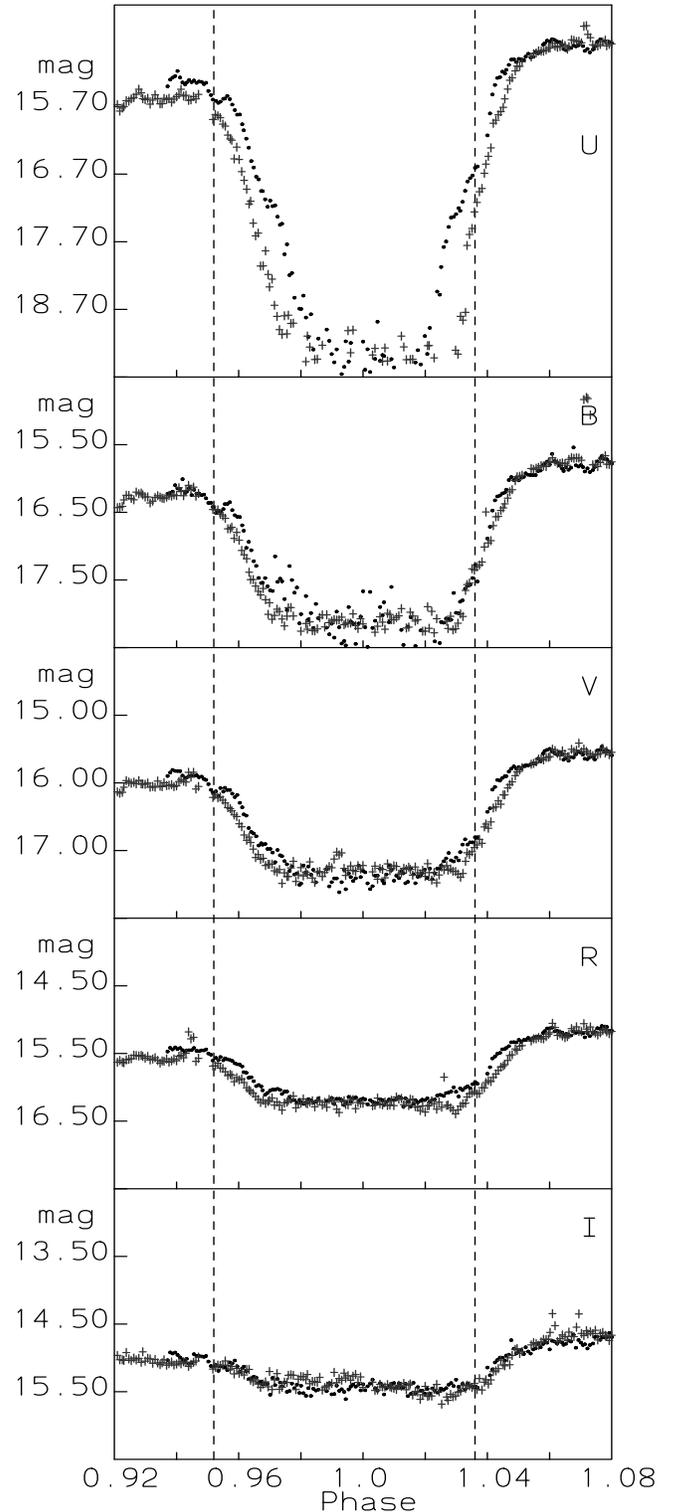,width=9.5cm,height=21cm}
 \caption{
V1309 Ori eclipse profiles ($UBVRI$), between ${\Phi=0.92 - 1.08}$.
Observations are plotted for two nights showing the largest difference in 
eclipse width: the 1997 Oct 6/7 (black circles) with the narrowest 
(most disturbed) 
eclipse is compared with the widest (flat bottom) light curve obtained on 
Nov 24/25 (crosses). Variations in the ingress and egress shape indicate 
different contributions from the extended accretion stream.
Dashed lines show the start and end of the white dwarf eclipse
according to Staude et al. \shortcite{b42}}.
\label{eclipse}
\end{figure}

The eclipse profiles of V1309 Ori are unusual compared to other
eclipsing AM Her stars such as HU Aqr or UZ For: the ingress and
egress of the eclipse are very shallow and have large night to night
variations (Figure \ref{eclipse}). This suggests that there is an
unusually prominent stream component. We have analyzed the colours of
the 'extra' emission defined as the flux difference between the
shallowest (most disturbed) eclipse and the widest (cleanest) flat
bottom light curve (Figure \ref{eclipse}). The resulting colour
indices for the additional flux $(U-B)_e$ = -1.0, $(B-V)_e$ = 0.1, and
$(V-R)_e$ = 0.5, match the (stream dominated) colours of V1309 Ori
outside the eclipse.  This supports the view that that eclipse shape
variations are caused by variations in the brightness or trajectory of
the accretion stream.

Since the eclipse is composed of an eclipse of the white dwarf and
stream, the mid-point of the eclipse may not be a good marker of the
inferior conjunction of the secondary. This is clearly showed by
Staude et al. \shortcite{b42}, who found that mid-eclipse of the
white dwarf occurs 172$\pm$20 s earlier than the observed mid-eclipse
seen in the light curves.

\begin{figure}
\psfig{file=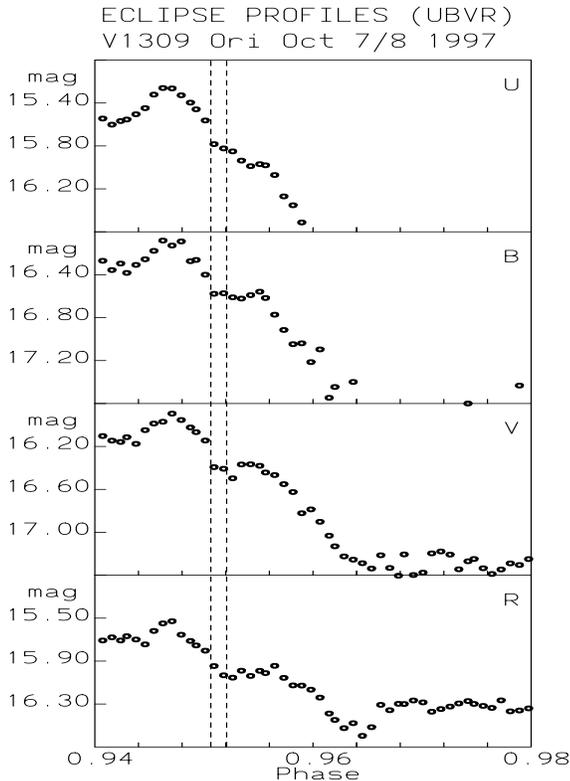,width=9cm,height=11cm}
  \caption{
The ingress in $UBVR$, observed at 
October 7/8, 1997, with the time resolution of 24 s. 
The two dashed lines shows the place of the 
ingress between phases ${\rm \Phi=0.951}$ ${\rm \Phi=0.952}$. This is the 
same phase as observed by Schmidt \& Stockman \shortcite{b33} and Staude et al.
\shortcite{b42} in HST far UV-data.}
\label{eclipse_detail}
\end{figure}

A more detailed examination of the eclipse profiles (Figure
\ref{eclipse_detail}) reveals a rapid eclipse ingress of the white
dwarf \& accretion region between ${\rm \Phi=0.951 - 0.952}$, seen for
the first time in the optical.  This drop is most prominent in data
obtained in 7/8 October 1997, while on the other observing nights in
October this phenomenon was not as prominent. With a time-resolution
of 24 sec, we cannot determine if this rapid drop is due to the
eclipse of the hot spot, or the whole white dwarf. The corresponding
egress of the white dwarf should take place at ${\Phi}$=1.036
according Staude et al. \shortcite{b42}. Our data (eg Figure 5),
hints at a rise at orbital phase 1.04, although it is far from clear.

Interestingly, time resolved HST UV-spectroscopy by Schmidt \&
Stockman \shortcite{b33} revealed a sharp drop in brightness, with duration of
6$\pm$2 sec at orbital phase $\Phi=0.952$, which they identified as
the eclipse ingress of hot accretion spot on the surface of the white
dwarf. Staude et al. \shortcite{b42} did not detect the ingress
and egress of the white dwarf in their broad band optical photometry,
but they re-analyzed the same HST-data on V1309 Ori and concluded that
the ingress (duration of 45$\pm$30 sec) which occurs at orbital phase
$\Phi=0.952$ is actually due to the eclipse of the whole white dwarf.

The observed brightness of V1309 Ori during the mid-eclipse in the
$V$-band is 17.3 magnitudes and colours: $U-B=1.3$, $B-V=0.9$,
$V-R=1.1$, and $R-I=0.7$.  The observed $B-V$ value corresponds to
that of a K2 dwarf, whereas other colours ($U-B$, $V- R$, $R-I$) fit
better K5, K7 or M0 dwarf \cite{b6}.  The orbital period - secondary
star relation given by Smith \& Dhillon \shortcite{b40} gives a much earlier
spectral type for the secondary (K2-K3). The observed $B-V$=0.9 shows
that there is $\sim$50 per cent more flux in the B-band compared to
that expected from a K7 spectral type dwarf. This extra flux might be
emitted by an accretion stream which was not completely eclipsed even
during the mid-eclipse.  However, HST-spectroscopy analyzed by Staude et al. 
\shortcite{b42}, showed that the eclipse of the stream and
the white dwarf was total in UV during their observations, implying
that this was unlikely.

\subsection{Quasi Period Oscillations}
 
We have studied this phenomenon by performing wavelet time-series
analysis of our photometric data. The main difference of using this
method instead of standard Fourier analysis is that the shape of the
(mother) wavelet can easily be adjusted to better correspond to the
shape of the flares in flickering (for example, more triangular).

The moderate time-resolution of 24 sec of photometry does not allow us
to search for variations on timescales much less than 1 min. CASLEO
observations (January 1998) are not included in this analysis due to
their higher noise level.  In our analysis we used Morelet wavelets
(for more details, see Lehto et al. 1999). Flickering up to 0.2
magnitudes occurs in the V1309 Ori light curves from night to
night. This is most prominent between the orbital phases
$\Phi=0.20-0.25$.

\begin{figure}
\psfig{file=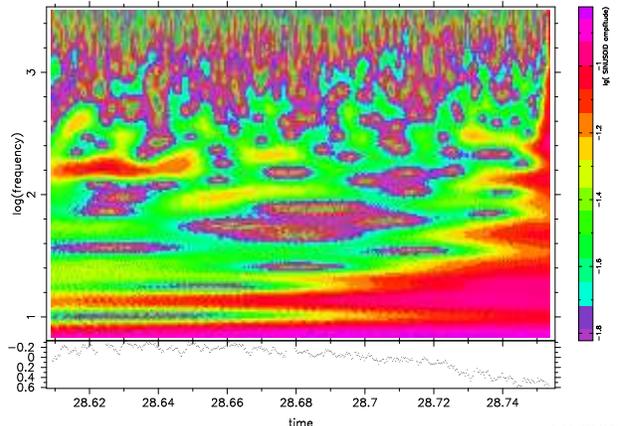,width=6cm,height=6cm}
\caption{V1309 Ori Morelet wavelet transform for U-band data, obtained on 
 6/7 October 1997.
The abscissa is linear time and the ordinate is the log of the inverse of
the timescale, where {\it f}=2 corresponds to P=1/100 days, i.e. 14.4 minutes,
and {\it f}=3 corresponds to P=1.44 minutes. A signal between 
28.62 (=HJD 2,450,728.62 - 2,450,700) and 28.65 corresponds variability 
timescales of 10 minutes.}
\label{qpo}
\end{figure}

Our results show evidence for variability on timescales of $\sim$10
minutes (NOT observations on 6/7 October 1997; Figure \ref{qpo}),
mostly in $UBV$. Other nights also similar variability. The QPOs and
flickering in the light curves have largest amplitudes near orbital
phases ($\Phi$=0.20-0.25) when the accreting pole emitting positive
circular polarization is seen face-on.

\section{Polarimetry}
 
\subsection{Circular polarization}

Our $UBVRI$ circular polarimetry, obtained on 24/25 November 1997,
over one complete orbital cycle (Fig. 2) shows both negative and
positive polarization. Positive circular polarization is observed
between phases $\Phi$=0.0 and $\Phi$=0.4, while negative
circular polarization is observed between phases $\Phi$=0.5 and
$\Phi$=0.9. The circular polarization show strong colour dependence:
polarization is strongest in $B$ and $V$, and is fairly modest in $U$
and $I$. This can be explained by the combined effects from cyclotron
emission and dilution by the stream.

The circular polarization variations indicate that accretion occurs
onto two separate regions. A single accreting pole can only give brief
sign reversal when the accreting region is near the limb of the white
dwarf, if the geometry is such that the angle between the line of the
sight and the magnetic field can go through 90$^{\circ}$. 

The circular polarization sign is reversed at the same orbital phases
(at $\Phi$=0.0 and $\Phi$=0.4) both in our October and November 1997
data, setting tight constrains on the degree of synchronism of the
spin of the white dwarf and the binary orbital period. This high
degree of synchronism is strengthened when we compare our observations
to the white light circular polarization data taken in December 1994
and January 1995 by Buckley \& Shafter \shortcite{b4}: the sign of the
circular polarization changes at the same orbital phases ($\Phi=0.0$
and $\Phi=0.4$) in their observations.  An uncertainty of about 0.05
of the phase difference of the zero crossovers over more than 3000
orbital cycles corresponds to the white dwarf spin and orbital periods
being equal to better than $\sim\times 10^{-5}$.

\begin{figure*}
\begin{center}
\setlength{\unitlength}{1cm}
\begin{picture}(12,6)
\put(-6.5,7.){\includegraphics{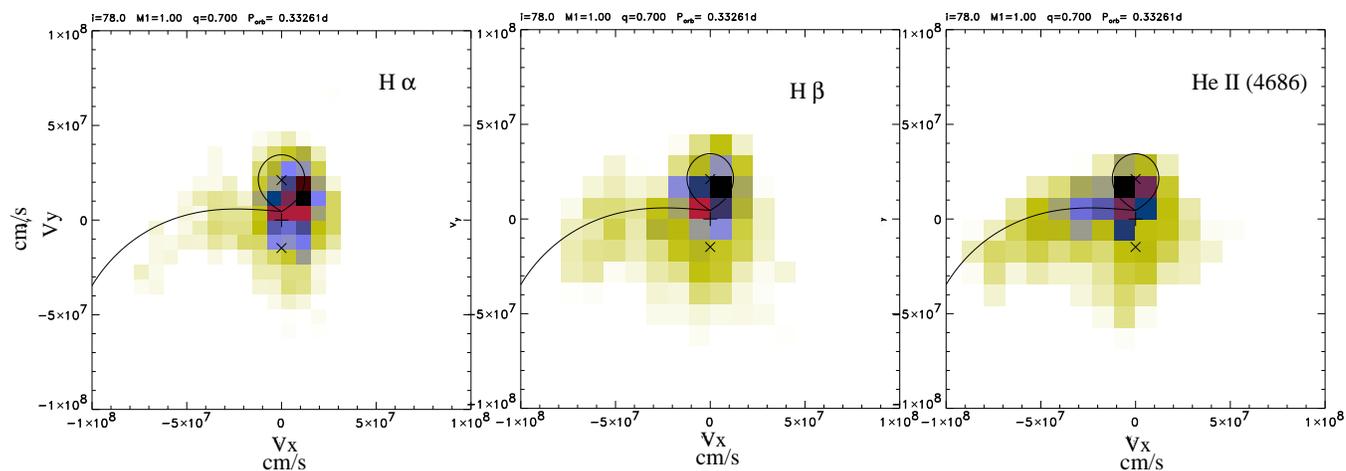}}
\end{picture}
\end{center}
\caption{Doppler tomograms of V1309 Ori in the lines of H$\alpha$,
H$\beta$ and HeII (4686).
The solid curve shows the ballistic trajectory of the accretion stream
and Roche lobe of the secondary.
The velocity of the primary and secondary is shown by a (x) and the center
of mass by a (+). The velocities are in units of cm/s.}
\label{doppler} 
\end{figure*}

\subsection{Linear polarization}

Figure \ref{linpol} (upper panel) combines our linear polarization
data obtained in October, November, 1997 (NOT) and in January 1998
(CASLEO).  These data have been vectorially averaged from individual
observations in the $UBVRI$ bands into 40 phase bins, in order to
increase the signal to noise ratio. Our linear polarization data cover
a complete orbital cycle, in contrast to previous studies which were
incomplete (Shafter et al. 1995; Buckley \& Shafter 1995). However,
the observed linear polarization in V1309 Ori is very low, less than
0.3-1.0 per cent during the orbital cycle, with no evidence for peaks
near the orbital phases where the circular polarization changes its
sign. Consequently, the position angle behaviour (Figure \ref{linpol},
lower panel) is noisy over the whole orbital cycle.

\section{Spectroscopy}
\subsection{Doppler tomography}

Doppler tomography has become a standard tool in the analysis of
interacting binary star research. Given a high enough spectral and
binary orbital resolution, a map can be made of the emission from a
line in velocity space. The first such map of a polar was made of VV
Pup \cite{b9}. Since then the technique has 
been applied to several other polars (eg Schwope et al. 1999). 
Here we obtain emission velocity maps of V1309 Ori using the code of 
Spruit (1998).

Since our spectra of V1309 Ori cover 0.7 of an orbital cycle rather
than a full cycle, we first determined if we could obtain useful
tomograms using spectra sampling an incomplete binary orbital
cycle. We extracted spectra of the polar HU Aqr from the data archive
of the Isaac Newton Group, La Palma. When we use only 0.7 of the
orbital cycle we find that the resulting Doppler tomograms are
remarkably similar to that when we use the full orbital cycle. There
are, however, some circular artifacts present in the maps and the
leading face of the secondary appears to be more strongly irradiated
rather than the trailing compared to the maps made using the full
dataset. This test indicates that we can expect to obtain meaningful
tomograms using spectra which do not fully sample the orbital cycle of
V1309 Ori. Indeed, Marsh \& Horne \shortcite{b22} showed using 
simulations that useful Doppler maps can be made even when the orbital 
cycle has been under-sampled. However, care should be taken in not 
over interpreting the details of those tomograms.

V1309 Ori was found to be in a high accretion state in December 1998
with the spectra showing prominent emission lines of H and He. We
generated Doppler tomograms of V1309 Ori in three emission lines
(H$\alpha$, H$\beta$ and He II $\lambda$4686) where we assume
$i=78^{\circ}$ and $q$=0.7 Staude et al. \shortcite{b42}. We show
in Figure \ref{doppler} the resulting tomograms.  Emission is seen
from the heated face of the secondary star and from the ballistic
component of the accretion stream. In the case of H$\alpha$, the bulk
of the line emission originates from the secondary star with the
stream emission being weaker. In He II the relative strength of these
two components is more equal.

Doppler maps of V1309 Ori have been presented by Shafter et al. 
\shortcite{b35} (although the spectral resolution was rather low), 
Hoard \shortcite{b16} who
presents maps in H$\beta$ and He II $\lambda$4686 (they cover 0.8 of
the orbital cycle) and also Staude et al. \shortcite{b42} 
who show maps in H$\gamma$, He II $\lambda$4686, 8236 and He I 4471 (they cover
the complete orbit). In the case of the He II map of Staude et al. 
\shortcite{b42} the
secondary and the stream are approximately equal, while in our map the
secondary is stronger and in the map of Hoard, emission from the
heated face of the secondary is much weaker than the stream.  Since
our maps do not sample the phase range between $\phi$=0.71--0.92, this
could be the reason why the secondary appears brighter.

\section{Archive X-ray data}

{\sl ROSAT} was an imaging X-ray satellite launched in 1990 with an
energy range $\sim$0.1-2.0keV. It had two X-ray instruments, the PSPC
and the HRI. The {\sl ROSAT} data of Walter, Wolk \& Adams \shortcite{b45}
showed evidence for a 'two pole' nature: during half of the orbital
cycle stronger X-ray emission is seen, and during the rest of the
cycle weaker emission is observed. We have extracted further {\sl
ROSAT} archive data of V1309 Ori and phased the data of Staude et al. 
\shortcite{b42}. We show in Table \ref{rosatdata} the data which we have
used. (There are three datasets in the archive where V1309 Ori was not
significantly detected).  We have scaled the data so that PSPC had an
effective area of 8 times that of the HRI.

\begin{figure}
\setlength{\unitlength}{1cm}
\begin{picture}(8,6)
\put(-0,-4.2){\includegraphics{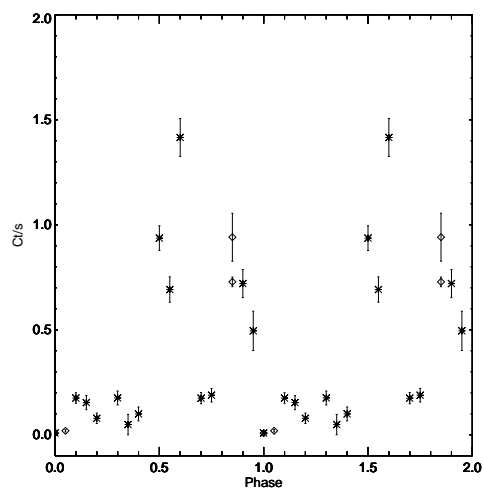}}
\end{picture}
\caption{Archive {\sl ROSAT} data of V1309 Ori folded on the ephemeris
of Staude et al. \shortcite{b42}. The diamonds represent data taken using the
PSPC and the stars the HRI.}
\label{rosat}
\end{figure}

We corrected the time of each event to the barycenter of the solar
system. We then phased each event on the ephemeris of Staude et al
(2001) and binned the light curve. (The ephemeris is accurate enough
that the phasing error is very small). We show this light curve in
Figure \ref{rosat}. There is some uncertainty in comparing these
various light curves since we do not know if they were in the same
accretion state. However, as found by Walter et al. \shortcite{b45}, V1309 Ori
is X-ray bright for approximately half the orbital cycle. It is faint
from $\phi\sim$0.0-0.5, and bright around $\phi\sim$0.5-0.6 and
0.8--0.95. The X-ray bright phases correspond to the phases which are
negatively circularly polarised.

\begin{table}
\begin{tabular}{llr}
\hline
Date & Instrument & Exp time (s)\\
\hline
1993 Sep 10 & PSPC & 3054 \\
1994 Feb 28 & PSPC & 1759 \\
1995 Sep 01 & HRI & 5825 \\
1996 Aug 25 & HRI & 14747\\
\hline
\end{tabular}
\caption{The observation log of the {\sl ROSAT} data. We show the
observation date, the instrument that was used and the effective
exposure.}
\label{rosatdata}
\end{table}

\section{Cyclotron model calculations}

We have modelled the observed $UBVRI$ circular polarization variations
(Figure \ref{circpol}) using simulation codes for cyclotron emission
described in Piirola et al. (1987a, b; 1990, 1993). The cyclotron
fluxes and polarization dependence on the viewing angle $\alpha$ (the
angle between the line of the sight and the magnetic field) have been
adopted from the model grids of Wickramasinghe \& Meggitt \shortcite{b46} for
constant temperature shocks (T$_{shock} = 10 - 40$ keV). These were
originally presented in 10 degree divisions, which we have then
interpolated in one degrees divisions over whole range of viewing
angle $\alpha$.  Calculations are then made by dividing emission
regions to equidistant points along a line on the white dwarf surface,
for which Stokes parameters Q, U, V and I are calculated independently
as an approximation for flat and extended regions. The height of the
shock is assumed to be negligible compared with the white dwarf
radius.

\subsection{Unpolarized background}

Based on the earlier studies of V1309 Ori (Shafter et al. 1995;
Schmidt \& Stockman 2001; Staude et al. 2001) and our
observations (\S 3) it is clear that variations in the light curves
are mostly caused by prominent stream emission with a smaller fraction
due to cyclotron emission. It is therefore necessary to include a
large unpolarized background when modelling the polarised light
curves. We have done this in two different ways: (1) with constant
background as first approximation, and (2) adopting the observed total
flux as a function of the orbital phase as the variable unpolarized
background.  The latter case is more realistic for V1309 Ori due to
the large amount of the unpolarized flux from the stream emission
which varies during orbital cycle. The diluting flux from stream
emission in V1309 Ori is 5--10 times larger than the peak cyclotron
flux, estimated from the fast drop of intensity at the eclipse of the
compact source observed at phase $\Phi=0.952$ (\S 3.1).  The
X-ray temperature is chosen ${\rm kT_{brems}}$= 10 keV according to
observations of de Martino et al. \shortcite{b9}.

\subsection{Parameters for accretion geometry}

Two-pole accretors, for example VV Pup \cite{b47}, UZ For \cite{b37}, 
DP Leo \cite{b7} and QS Tel \cite{b38},
have been found to show different magnetic field strengths in the
accretion regions located in opposite hemispheres: up to factor of 2
difference in the field strengths has been measured. The more strongly
accreting pole has normally the weaker magnetic field.  In the case of
V1309 Ori there are no major differences seen in the wavelength
dependence of the positive and negative excursions of the observed
circular polarization curves (Figure \ref{circpol}), suggesting that
both accreting regions are accreting nearly equally.  Therefore, we
have chosen the electron temperature and the plasma parameter
($\Lambda$ =10$^{5}$) to be the same in both regions. We assume an
inclination of {\it i}=78$^{\circ}$ Staude et al. \shortcite{b42}.
Parameters such as the longitude of the emission region(s) on the
surface of the white dwarf, and extension of the accretion regions
were varied to match the gross features seen in the circular
polarization behaviour. Due to very low level of the linear
polarization (less than 0.5 per cent), and noisy position angle
variations, we have not tried to use linear polarization to fix any
model parameters. Another reason for not doing so is that scattering
from free electrons of the stream can introduce significant linear
polarization effects, dominating over the low linearly polarized flux
of cyclotron origin.

Estimates of the colatitude of the accretion region $\beta$ can be
made if we assume that the observed circular polarization from
different poles is not significantly affected by possible overlap of
the polarized emission from another accretion region. We can estimate
$\beta$ for this region using the duration of the self-eclipse of the
accretion region and equation (1) of Visvanathan \& Wickramasinghe
\shortcite{b44}. Using the circular polarisation data shown in Figure
\ref{circpol}, we find ${\beta=145^\circ}$ for the positive pole, and
${\beta=35^\circ}$ for the negative pole, assuming an inclination
${\it i=\rm 78^{\circ}}$ by Staude et al. \shortcite{b42}.

\subsection{Results from the modelling}
\label{modelling}

Observed circular polarization variations can be reproduced reasonably
well with a model consisting of two separate emission regions, one
centered at colatitude $\beta$ = 145$^{\circ}$ (the positive pole),
seen closest to the observer at ${\Phi=0.20}$, and another region
centered in at $\beta$ = 35$^{\circ}$ (the negative pole), seen
closest to observer ${\Phi=0.70}$. For both regions we have adopted in 
Figure 10 longitudinal extension of 30$^{\circ}$ (in white dwarf 
rotational coordinates), but these values are not strongly constrained.
Extensions in the range 10$^{\circ}$ -- 60$^{\circ}$ give almost similar 
results. Point-like emission region gives too sharp polarization 
variations and very extended emission regions (larger than $60^{\circ}$) 
too smooth and low-amplitude curves.

The model shown in Figure {\ref{polmodel} assumes that cyclotron
harmonics 6, 5, 4, 3, and 2, dominate in the $UBVRI$ passbands,
respectively.  This corresponds to a magnetic field of about 50 MG,
which is similar to the estimated values for magnetic field (33-- 55
MG Garnavich et al. 1994; 61 MG Shafter et al. 1995). For a 50 MG
field the wavelengths of the harmonics 6 to 3 are at ${\rm 3580~\AA}$,
${\rm 4300~\AA}$, ${\rm 5370~\AA}$ and ${\rm 7160~\AA}$, i.e. one
cyclotron harmonic clearly dominates in each of the $UBVR$ bands. Our
$I$ band (${\rm 8300~\AA}$) falls about half-way between the 3rd and
2nd harmonics at ${\rm B\sim50~MG}$.  The best correspondence to the
observed circular polarization variations is achieved using the model
where unpolarized background varies in a similar way to the total
observed flux over the orbital cycle (Figure \ref{polmodel}, continuous
line).

Our model parameters for the location of accretion regions (in white
dwarf rotation coordinates), positive pole at ${\rm \beta =
145^{\circ}, \Psi=-70^{\circ}}$ and negative pole at ${\rm \beta =
35^{\circ}, \Psi=110^{\circ}}$, are similar to the values reported by
Harrop-Allin et al. \shortcite{b15}, who modelled white light data of Buckley
\& Shafter \shortcite{b4}: $\beta = 40^{\circ}$ and $\beta = 140^{\circ}$. In
contrast Staude et al. \shortcite{b42} derived values of $\beta$ =
17$^{\circ}$ and $\Psi=-16^{\circ}$ from their Doppler maps. In that
study it was assumed that only one accretion region was
visible. Although it is possible for an accretion region to show both
positive and negative circular polarisation, it is only for a short
phase duration if we observe the `underside' of the shock.

\begin{figure}
\psfig{file=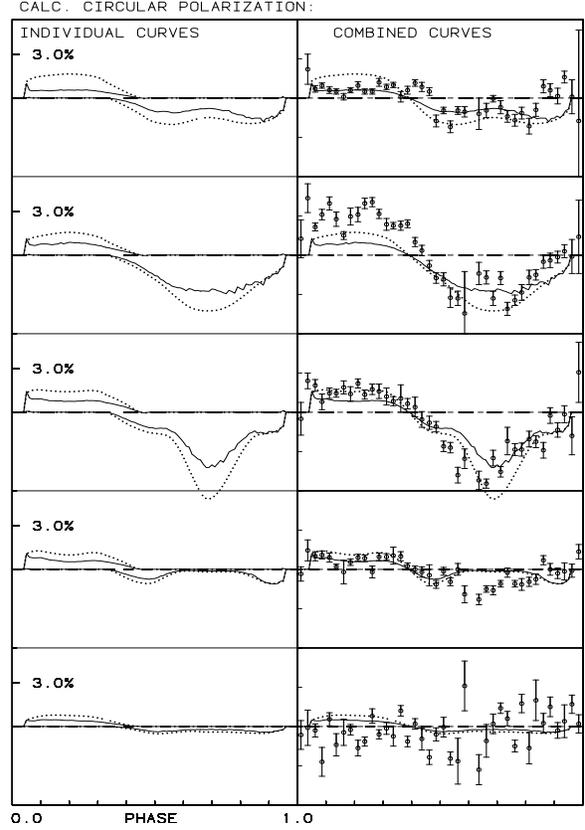,width=8.5cm,height=12cm}
\caption{ Calculated circular polarization curves for cyclotron
emission harmonics from 6th (top) to 2nd (bottom) from cyclotron
models with ${\rm kT = 10keV}$, ${\it \Lambda} = 10^{5}$, {\it i}
=78$^{\circ}$ and two extended emission regions separated by about 0.5
orbital phase, and centered at colatitudes $\beta=35^{\circ}$
(negative pole) and $\beta= 145^{\circ}$ (positive pole).  In the left
panel emission from each of the separate emission regions are
presented and in the right panel the combined curves are presented
together with the observed data from $UBVRI$ bands.  The dotted line
shows the polarization from cyclotron emission with constant
background, and the continuous line shows the case where total flux
and its orbital variations is adopted as the unpolarized diluting
background.}
\label{polmodel}
\end{figure}

\section{Discussion}

\subsection{A well synchronised system}

By comparing our circular polarization curves with those found in the
literature we confirm that the spin of the white dwarf in V1309 Ori is
synchronised with the orbital period to a high degree. The
zero-crossings of circular polarization take place at the same phase
of the orbital period as found by Buckley \& Shafter \shortcite{b4}, which
puts an upper limit of ${\rm \sim0.002}$ per cent for the difference
of the white dwarf spin and orbital period. There are four polars
which have been found to show a small ($\sim$1 percent) degree of
asynchronism. The polar showing the smallest, V1432 Aql, is 0.28
percent asynchronous \cite{b13}: over two orders of
magnitude greater than V1309 Ori.

\subsection{Accretion geometry}

Our circular polarimetry curves (Figure 2) shows clear positive and
negative excursions which indicate that V1309 Ori has two accreting
poles. This is consistent with previous polarisation observations.
The X-ray data from the {\sl ROSAT} archive suggests that the negative
circularly polarised pole is brighter in X-rays. It is possible that
it is bright because of the increased mass transfer at this
pole. Alternatively, it might be due to the fact that the accretion
flow to this pole is very inhomogeneous, with the dense parts of the
flow accreting directly into the white dwarf without causing a shock
and therefore liberating its energy at soft X-rays.

We note that Staude et al. \shortcite{b42} based on optical/UV photometry and
optical spectroscopy did not find any evidence for a second accretion
pole, although they could not exclude one. Their modelling predicts
that the one accreting pole would show a maximum in the soft X-ray
light curve at $\Phi$=0.045 and would show a self eclipse at
$\Phi$=0.55. However, the {\sl ROSAT} data does not confirm this
view. Indeed, we see maximum flux at $\phi$=0.55 and a minimum at
$\Phi$=0.045.

The X-ray data does show a dip around $\Phi\sim$=0.7. This is unlikely
to be due to absorption of X-rays by the accretion stream since the
stream does not cut through our line of sight to the accretion region
at this phase. It is also unlikely that this dip is due to that
observation being at a lower accretion state since the same
observation shows a peak at $\phi\sim$=0.6. The cause of the dip in
X-rays is unclear but maybe due to a fraction of the accretion region
being self-eclipsed by the white dwarf at these phases.

\subsection{The mass of the white dwarf}
\label{mass}

Along the ballistic flow, emission extends to ${\rm V_{x}\sim-800}$ km
s$^{-1}$, ${\rm V_{y}\sim-200}$ km s$^{-1}$ in our maps and also the
maps of Hoard (1999). The maps of Staude et al.\shortcite{b42} show the flow
extending to ${\rm V_{y}\sim-100}$ km s$^{-1}$. Taking $V_{x}, V_{y}$
from our maps (giving a velocity of 820 km s$^{-1}$) and using
$v=(2GM_{wd}/r)^{1/2}$ (where $v$ is velocity and $r$ is distance from
the white dwarf), we find that the end of the ballistic flow is
$r$=2.4 and $2.8\times10^{10}$ cm distant from the white dwarf for
$M_{wd}$=0.6 and 0.7\Msun respectively.

We can compare these estimates to the expected distance from the white
dwarf that material gets coupled by the magnetic field ($R_{\mu}$)
using equation (1b) of Mukai \shortcite{b23}. We assume $B$=50 MG (typical of
the estimates made for V1309 Ori), $\sigma_{9}$=3 (the radius of the
stream in units of $10^{9}$ cm: the value estimated for HU Aqr,
Harrop-Allin et al. 1999) and $\dot{M}_{16}$=10 (the mass transfer rate
in units of 10$^{16}$ g s$^{-1}$, Harrop-Allin et al. 1997). We find
that for $M_{wd}$=0.6 and 0.7\Msun we get $R_{\mu}$= 3.4 and
2.1$\times10^{10}$ cm respectively. Although there is some
considerable degree of uncertainty in how applicable the above
formulation for $R_{\mu}$ actually is, it is interesting that for
masses between 0.6--0.7\Msun the predicted value of $R_{\mu}$ is
consistent with our Doppler maps.  This range of mass is consistent
with that estimated by Staude et al. \shortcite{b42}.

\subsection{QPOs and their timescales}

Our photometric observations show evidence for QPOs on time scales of
10 min with amplitudes up to 0.2 magnitudes. This compares with 6.7
and 15.5 min: Shafter et al. \shortcite{b35}. There are some examples of other
AM Her systems where flickering on time scales of few minutes have
been observed: 4.5 min in BL Hyi \cite{b39}, 4 -- 11 min in
QQ Vul \cite{b24}, and in AI Tri 6.5-- 7 min and 13.5 -- 14
min \cite{b36}.  We speculate that the QPOs in V1309 Ori 
are due to 'blobby accretion', already observed in X-ray data 
(Walter, Wolk \& Adams 1995; de Martino et al. 1998).

We compare these timescales with those derived by King \shortcite{b17}, and
King \shortcite{b18} (see also Chanmugam 1995). The irradiation of the
accretion flow may ionise the subsonic accretion flow below the inner
$L_{1}$ point and modulate gas flow through this point on the
timescale of the dynamical time scale in the Roche potential near
$L_{1}$.  The equation presented by King \shortcite{b17}

\begin{equation}
~~~~~~~~~~~~~~~~~T_{osc}\sim\frac{H_{*}}{c_{*}}=5.5\times10^{-2}P
\end{equation}

where ${\rm H_{*}}$ is the scale height and ${\rm c_{*}}$ is local
sound speed near ${\rm L_{1}}$, predicts the timescales for these
oscillations. The orbital period of 479 minutes gives us a prediction
of 26 minutes for V1309 Ori. The observed timescales of QPOs in V1309
Ori, 10 and 15 minutes, are approximately half of the predicted. The
QPOs are seen strongest in $UBV$, and are negligible in the longer
wavelengths, which may due to reason that only a fraction of flow will
undergo these oscillations near ${\rm L_{1}}$, as pointed out by King
\shortcite{b18}.

\subsection{Eclipse profiles}

Several eclipsing polars have been observed with high signal to noise
and high time resolution. These include HU Aqr \cite{b15} and UZ For 
\cite{b26}. Both of these systems
show a sharp eclipse ingress lasting several seconds. This sharp drop
in intensity is associated with the eclipse of the bright accretion
region on the white dwarf. In the case of UZ For there are two sharp
intensity drops, indicating that there are two accretion regions
visible during the eclipse.

One of the most striking features about the eclipse profiles of V1309
Ori is the obvious lack of a sharp ingress or egress which indicate
the (dis)appearance of the white dwarf and/or hot spots in the surface
of the white dwarf behind the secondary. The relative faintness of the
accretion region(s), compared to the bright accretion stream in V1309
Ori, may be the reason for this.

After the eclipse of the white dwarf, the accretion stream is still
visible for a length of time. Observations of other polars such as HU
Aqr (Glenn et al. 1994, Bridge et al. 2002) show that this length of
time can vary from one cycle to the next. The fact that we observe a
variable eclipse ingress of the accretion stream in V1309 Ori is
therefore not un-typical of polars. However, what does make V1309 Ori
unique amongst polars is the fact that the eclipse egress is highly
variable: all other polars show a rapid rise at the same phase coming
out from eclipse. The fact that V1309 Ori does not implies either that
we can observe the stream above the orbital plane before the white
dwarf is visible, or that the accretion stream travels far enough
around the white dwarf so that it is visible before the white dwarf
itself. 

To investigate if this further, we shown in Figure \ref{system} the
view of the system at two different phases ($\Phi$= 0.96 and 1.04). We
use the following system parameters in determining these:
$i=78^{\circ}$, $q$=0.67, $M_{wd}=0.7$ \Msun~\cite{b42} and a white 
dwarf - secondary star separation of
1.47$\times10^{11}$ cm (determined using the above parameters and
standard Roche lobe geometry). We also show a single magnetic field
line originating from the negative circularly polarised accretion
region ($\beta=35^{\circ}$ and face onto the observer at $\Phi$=0.2,
\S \ref{modelling}). In Figure \ref{system}, the accretion streams leading
to both poles are visible at $\rm \Phi=0.96$. At $\rm \Phi=0.04$
only the stream leading to the negative pole is visible. The white
dwarf appears before the stream leading to the positive pole is
visible. Even if the negative pole is not visible at 0.04, the emission
from the stream leading to that pole is.
We take $R_{\mu}/a$=0.2: this implies
$R_{\mu}=2.9\times10^{10}$ cm. This is consistent with our findings in
\S \ref{mass}. This shows that for this accretion stream geometry the
accretion stream is visible after the white dwarf has been eclipsed
and also before the white dwarf comes out of eclipse. If the stream
emission was highly variable then this could explain the variable
egress profile.

\begin{figure*}
\begin{center}
\setlength{\unitlength}{1cm}
\begin{picture}(8,6)
\put(-3,-2.){\includegraphics{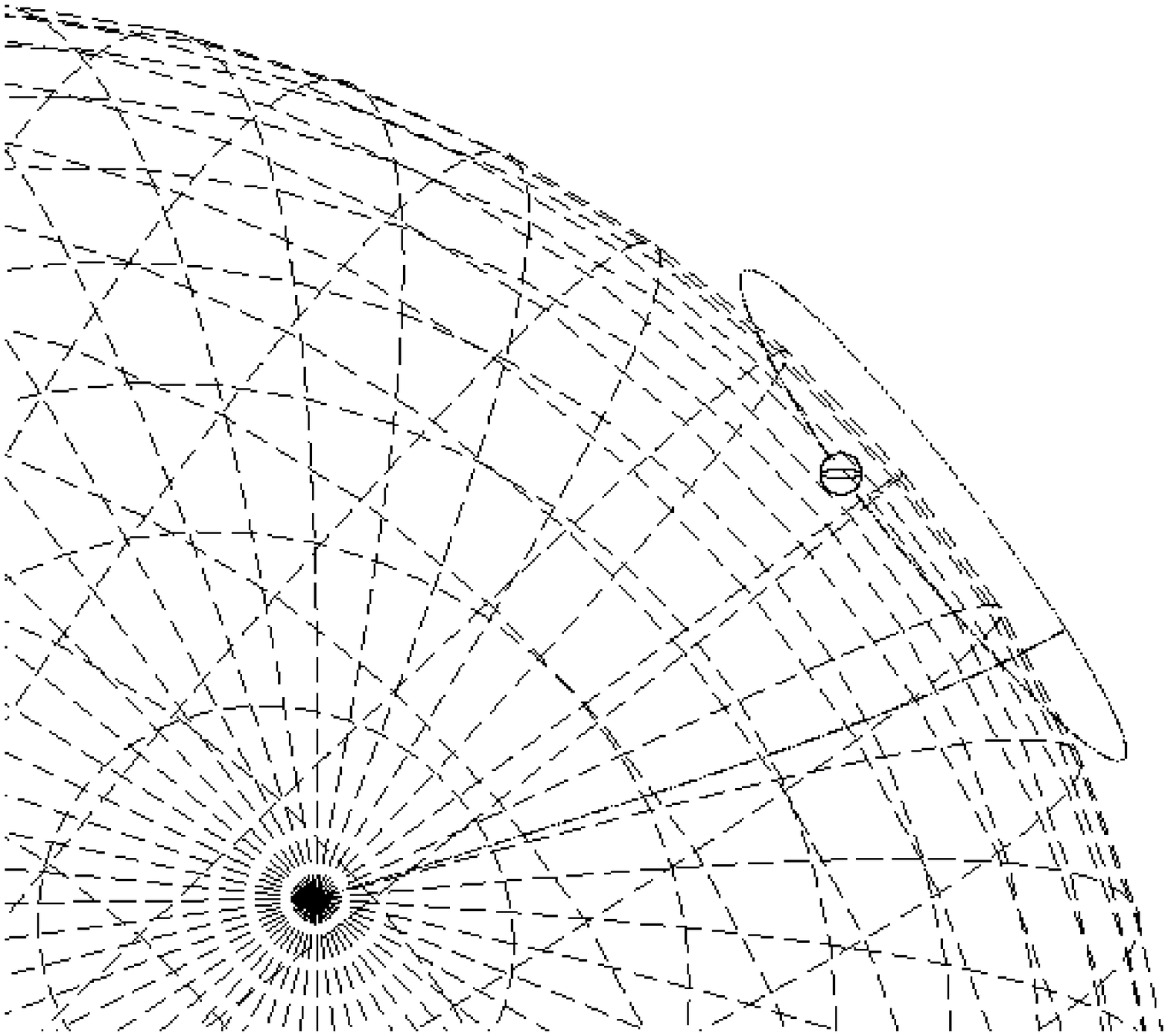}}
\put(3,-2.){\includegraphics{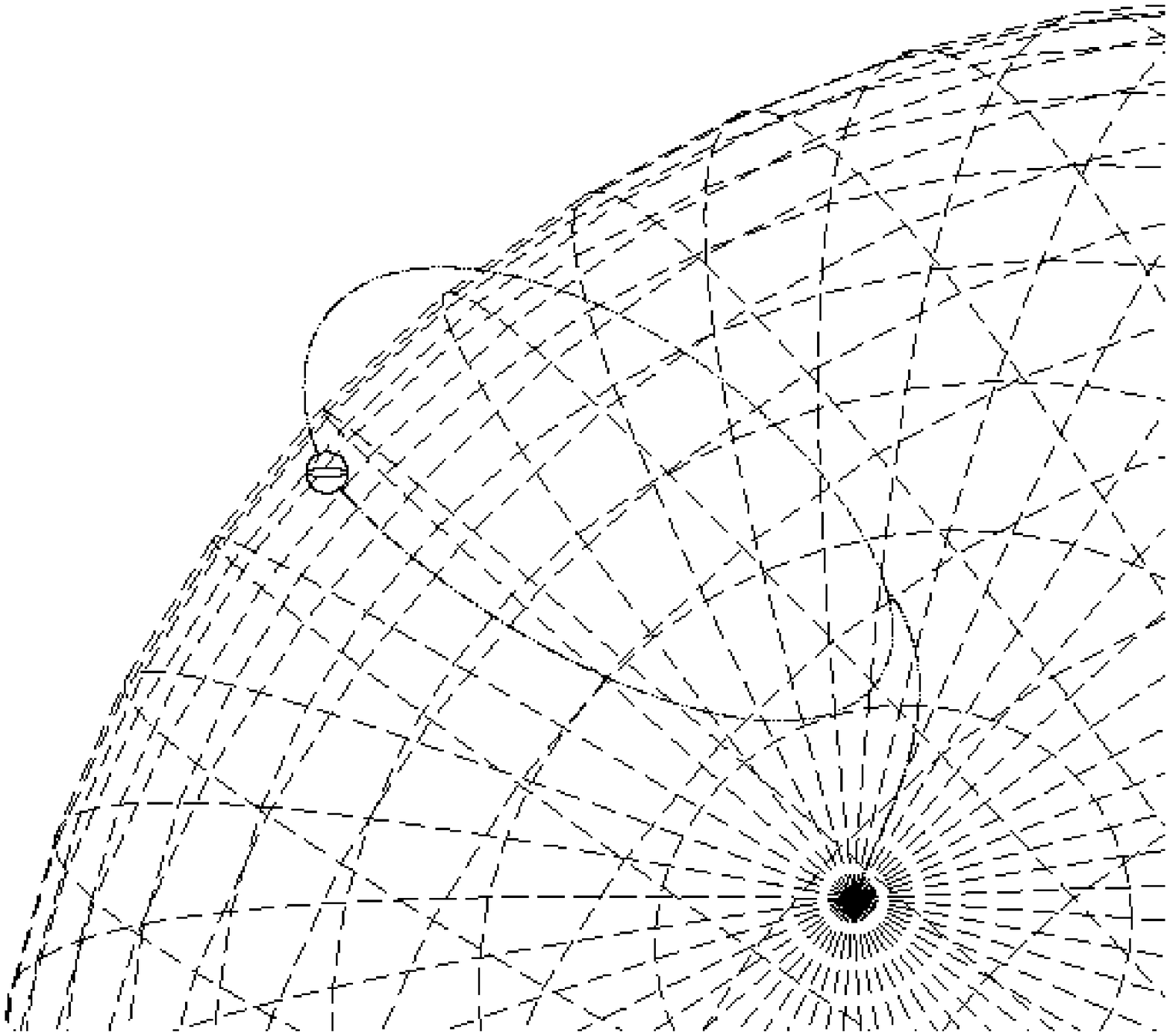}}
\end{picture}
\end{center}
\caption{The view of the system at phases $\Phi$=0.96 and 1.04. See
the text for details of the system parameters and accretion geometry
which are assumed. The secondary Roche lobe is projected as a wire
model so that the white dwarf and stream remain visible.}
\label{system} 
\end{figure*}

\subsection{Evolution}
          
Garnavich et al. \shortcite{b12} and Shafter et al. \shortcite{b35} noted 
that the
secondary star in V1309 Ori is oversized for its spectral type (M0 -
M1) and mass (0.4--0.6 $M_{\odot}$). Indeed, recent binary evolution
models (eg Smith \& Dhillon 1998 and Baraffe \& Kolb 2000) which
assume an unevolved donor star and typical mass transfer rates, all
predict either a much earlier or later spectral type for V1309 Ori
than observed (cf Figure 1 of Baraffe \& Kolb 2000).

However, by assuming an evolved donor it is possible to match the
observed spectral type to the predicted value. For instance, from
Figure 3 of Baraffe \& Kolb \shortcite{b1}, for an initial secondary star mass
of $M_{2}$=1.2\Msun, a central Hydrogen abundance at the start of mass
transfer of $X_{c}$=0.05, and a mass transfer rate of $\dot{M} =1.5
\times 10^{-9} M_{\odot} (\sim1\times10^{17}$) g s$^{-1}$) we
find we that a spectral type of M1 is predicted for an orbital period
of 8 hrs. This is within the estimated range of values required to
satisfy conditions for observed oversized secondary to fill its Roche
lobe (the main uncertainties are the distance and the mass of the
white dwarf, eg  Harrop-Allin et al. 1997, de Martino et al. 1998).

As already noted by King, Osborne \& Schenker \shortcite{b19}, V1309 Ori is
indeed a good possible candidate for a binary system which has gone
through a supersoft source phase and contains a nuclear evolved donor
star. Szkody \& Silber \shortcite{b43} and Schmidt \& Stockman \shortcite{b33} 
have noticed that V1309 Ori has extraordinary strong excitation lines of ${
\rm N~V~\lambda 1240}$ and ${ \rm O~V~\lambda1370}$) which may support
such an interpretation.
\section{Acknowledgments}

Nordic Optical Telescope is operated on the island of La Palma jointly
by Denmark, Finland, Iceland, Norway, and Sweden, in the Spanish
Observatorio del Roque de los Muchachos of the Instituto de
Astrofis\'{\i}ca de Canarias.  This work has been supported by the
European Commission through the Activity ``Access to Large Scale
Facilities'' within the Programme ``Training and Mobility of
Researchers'' awarded to the Instituto de Astro\'{f}isica de Canarias
to fund European Astronomers access to its Roque de Los Muchachos and
Teide Observatories (European Northern Observatory), in the Canary
Islands.  SK is grateful for the Finnish Academy of Sciences and
Letters for support.  We would like to thank also the Australian
National University for the allocation of telescope time and Dr Henk
Spruit for the helpful advice and generous use of his Doppler
Tomography software. The {\sl ROSAT} data were extracted from the
Leicester Database and Archive Service.

\label{lastpage}
\end{document}